\documentclass[twocolumn,floatfix,superscriptaddress,pra,aps,10pt,longbibliography]{revtex4-2}
\usepackage{algorithm}
\usepackage{algpseudocode}
\usepackage{times}
\usepackage{amssymb}
\usepackage{color}
\usepackage{amsmath}
\usepackage{amsbsy}
\usepackage{amsthm}
\usepackage{graphicx}
\usepackage{bbm}
\usepackage{bm}
\usepackage{epsfig}
\usepackage{xfrac}
\usepackage{xcolor}
\usepackage{enumerate}
\usepackage{multirow}
\definecolor{myurlcolor}{rgb}{0,0,0.7}
\definecolor{myrefcolor}{rgb}{0.8,0,0}
\usepackage[unicode=true,pdfusetitle, bookmarks=true,bookmarksnumbered=false,bookmarksopen=false, breaklinks=false,pdfborder={0 0 0},backref=false,colorlinks=true, linkcolor=myrefcolor,citecolor=myurlcolor,urlcolor=myurlcolor]{hyperref}
\newcommand{\dd}{\mathrm{d}}
\newcommand{\ket}[1]{\left| {#1} \right\rangle}
\newcommand{\bra}[1]{\left\langle {#1}\right|}
\newcommand{\ketbra}[2]{\ket{#1}\!\bra{#2}}

\def\sectiontext#1{{\it #1.---}}
\def\>{\rangle}
\def\<{\langle}
\renewcommand{\t}[1]{\textrm{#1}}
\newcommand{\tr}[0]{\mathrm{Tr}}

\newcommand{\intinf}{\int_{-\infty}^{+\infty}}

\usepackage{braket}
\newcommand{\ip}[2]{\left\langle #1 \middle| #2 \right\rangle}
\newcommand{\dyad}[1]{\ket{#1}\bra{#1}}
\newcommand{\thmref}[1]{\hyperref[#1]{Theorem~\ref{#1}}}
\newcommand{\lemmaref}[1]{\hyperref[#1]{Lemma~\ref{#1}}}
\newcommand{\figref}[1]{\hyperref[#1]{Fig.~\ref{#1}}}
\newcommand{\figaref}[1]{\hyperref[#1]{Fig.~\ref{#1}a}}
\newcommand{\figbref}[1]{\hyperref[#1]{Fig.~\ref{#1}b}}
\newcommand{\figcref}[1]{\hyperref[#1]{Fig.~\ref{#1}c}}
\renewcommand{\eqref}[1]{\hyperref[#1]{Eq.~(\ref{#1})}}
\newcommand{\eqsref}[2]{\hyperref[#1]{Eqs.~(\ref{#1})-(\ref{#2})}}
\newcommand{\appref}[1]{\hyperref[#1]{Appx.~\ref{#1}}}

\newcommand{\var}{\phi}
\newcommand{\Var}{\Phi}
\newcommand{\X}{X}

\newcommand{\trho}{\tilde{\rho}}

\newcommand{\ddvar}{\frac{\dd}{\dd\var}}

\usepackage{dsfont}

\newtheorem{theorem}{Theorem}
\newtheorem{lemma}{Lemma}

\begin{document}
\title{Mutual Information Bounded by Fisher Information}
\author{Wojciech G{\'{o}}recki}
\thanks{These two authors contributed equally to the project.}
\affiliation{INFN Sez. Pavia, via Bassi 6, I-27100 Pavia, Italy}
\author{Xi Lu}
\thanks{These two authors contributed equally to the project.}
\affiliation{School of Mathematical Science, Zhejiang University, Hangzhou, 310027, China}
\author{Chiara Macchiavello}
\affiliation{INFN Sez. Pavia, via Bassi 6, I-27100 Pavia, Italy}
\affiliation{Dip. Fisica, University of Pavia, via Bassi 6, I-27100 Pavia, Italy}
\author{Lorenzo Maccone}
\affiliation{INFN Sez. Pavia, via Bassi 6, I-27100 Pavia, Italy}
\affiliation{Dip. Fisica, University of Pavia, via Bassi 6, I-27100 Pavia, Italy}

\begin{abstract}
  We derive a general upper bound to mutual information in terms of
  the Fisher information. The bound may be further used to derive a
  lower bound for the Bayesian quadratic cost. These two provide
  alternatives to other inequalities in the literature (e.g.~the van Trees inequality) that
  are useful also for cases where the latter
  ones give trivial bounds. We then generalize them to the quantum case, where they bound the Holevo information in terms of
  the quantum Fisher information. We illustrate the usefulness of our bounds
  with a case study in quantum phase estimation. Here, they allow us
  to adapt to mutual information (useful for global strategies where the prior plays an important role) the known and highly nontrivial
  bounds for the Fisher information in the presence of noise. The results are also useful in the context of quantum communication, both for continuous and discrete alphabets.
  \end{abstract}

\maketitle

The Fisher information (FI) $F(\var)$ measures how much information
a conditional probability distribution $p(x|\var)$ contains on some
parameter $\var$. Instead, the mutual information (MI)
$I(\X,\Var)$ measures how much information two random variables $\X$
and $\Var$ have on one another. Then, it is clear that the two
quantities must be related if one considers the parameter $\var$ as
realization of an unknown random variable $\Var$, and indeed many such relations have
appeared in the literature
\cite{brunel1998mutual,bethge2002optimal,yarrow2012fisher,wei2016mutual,czajkowski2017super,barnes2021fisher}. Yet,
most of them work only in the asymptotic limit and/or require
additional assumptions about the regularity of probability
distributions. The most relevant one is the Efroimovich
inequality~\cite{efroimovich1980information,aras2019family,lee2022new},
valid for a finite number of samples without additional
assumptions. It is a generalization of the van Trees inequality (or
Bayesian Cram\'er-Rao
bound)~\cite{schutzenberger1957generalization,van2004detection,gill1995}. However, both of them might
fail to give significant bounds, e.g.~when the prior probability on
$\var$ has sharp edges, such as the important case of a uniform prior
on a finite interval.

\begin{figure}[t!]
\includegraphics[width=0.48\textwidth]{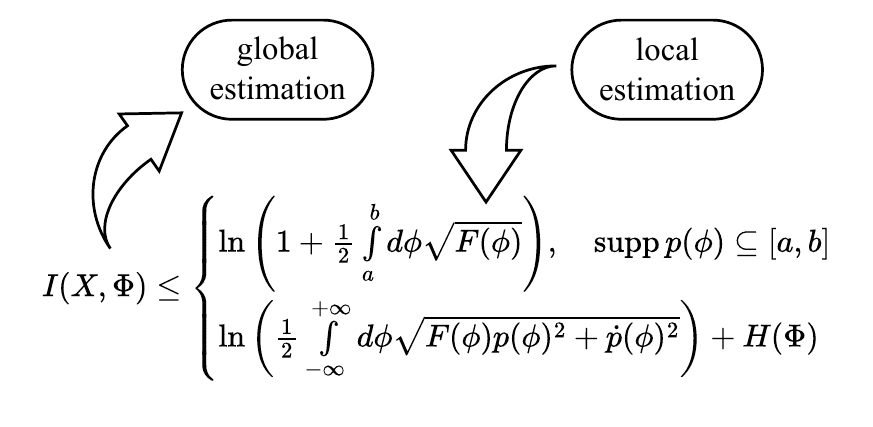}
\caption{The bound for the mutual information $I(\X,\Var)$ in terms of the Fisher information $F(\var)$ allows for the transfer of the results obtained for local estimation to global estimation. Here the parameter $\var$ is considered as realization of a random variable $\Var$.}\label{fig:abstr}
\end{figure}

Here we derive two universal upper bounds to the mutual information in
terms of the Fisher information which do not suffer from these issues:
a simple one, valid in the useful case where the prior distribution
$p(\var)$ for $\var$ has finite support, and a general one valid
for any prior. Our bounds provide a bridge between {\em local} and
{\em global} estimation (Fig.~\ref{fig:abstr}). For local estimation
when a large
number of probes is under consideration the FI is a sufficient tool for
meaningful analysis. If, instead, one needs to take into account also
a nontrivial prior information $p(\var)$, then {\em global}
estimation approaches, such as the one based on mutual information
\cite{hall2012does,hall2012universality,rzadkowski2017discrete,hassani2017digital,chesi2023general}
or the Bayesian
approach~\cite{Gorecki2020pi,morelli2021bayesian,gebhart2021bayesian,dinani2019bayesian},
or the minimax
cost~\cite{butucea2007minimax,hayashi2011,gorecki2021multiple} are
more useful. Since an upper bound on the entropy of a probability
distribution imposes a lower bound on its moments, our bound to MI
immediately implies the bounds for any Bayesian
cost~\cite{chen2024l_q}.  Therefore, the bounds for MI are more
meaningful than the bounds on Bayesian cost, in the same way as the
entropic uncertainty relations are stronger than the standard
ones~\cite{bialynicki1975uncertainty,hall1993phase,coles2017entropic}. The
most famous example is the relation between MI and averaged mean
square error~\cite{hall2012does,hall2012universality}. Using this
relationship, our result also sets a bound on the squared error,
operating for a broader class of problems than the standard van Trees
inequality.

In quantum mechanics, where conditional probability $p(x|\var)$ comes from performing measurements on a quantum state, MI maximized over the choice of measurements is known to be bounded by Holevo information~\cite{wilde2011classical}. Therefore, we also present the quantum version of the bound, which connects the Holevo information with the quantum Fisher information (QFI).

%We also show that these bounds imply a lower bound to the Bayesian quadratic cost, namely to the average mean square error, averaged over the prior $p(\var)$. Importantly, our bound is useful for the cases in which the \sisi{remarkable}  van Trees inequality~\cite{van2004detection,gill1995} fails to give significant bounds, e.g.~for the case in which the prior has sharp edges, such as the important case of a uniform prior on a finite interval.

To show the usefulness of our bounds, we apply them to a case study in
quantum metrology. Quantum metrology
\cite{giovannetti2006quantum,Paris2009,giovannetti2011advances,Toth2014,demkowicz2015optical,Schnabel2016,degen2017quantum,Pezze2018,Pirandola2018}
is the study of how quantum effects such as entanglement
\cite{giovannetti2006quantum} or squeezing \cite{maccone2020squeezing}
can be used to enhance the precision of the measurement of a parameter
$\var$ of a quantum system. In the noiseless case, it is easy to
show that the ultimate limit in precision when the system is sampled
$N$ times, the Heisenberg scaling $\Delta\var\propto 1/N$, can be
achieved only using quantum effects, except in the trivial case in
which a single probe samples the system repeatedly and is measured
only once. Instead, classical strategies are limited to the standard
quantum limit (SQL) $\Delta\var\propto1/\sqrt{N}$ of the central
limit theorem. In the noisy
case~\cite{fujiwara2008fibre,kolodynski2010phase,escher2011general,demkowicz2012elusive,knysh2014true,demkowicz2014using,demkowicz2017adaptive,zhou2021asymptotic,kurdzialek2022using}
the situation is much more complicated, and, even though typically in
the asymptotic regime of large $N$ the Heisenberg scaling cannot be
achieved anymore, one can still obtain quantum enhancements by large
factors. Most of these results have been obtained using Fisher
information techniques, such as the quantum Cramer-Rao bound
~\cite{fujiwara2008fibre,escher2011general,demkowicz2012elusive,knysh2014true,demkowicz2014using,demkowicz2017adaptive,zhou2021asymptotic,kurdzialek2022using}. As
we show below, our new bounds allow us to transfer the highly
sophisticated results on FI in the presence of noise to the MI in a
simple way. In the noisy case, this provides a nontrivial bridge
between local and global quantum estimation strategies.
Besides quantum metrology, our result may be used to bound channel capacity in quantum communication.

%In the noiseless case, instead, these bounds can only be used to exclude the attainability of the Heisenberg scaling for separable strategies. So to find and optimize the strategies that do achieve the Heisenberg scaling for MI, we resort to a different analysis.

\sectiontext{Bounding mutual information by Fisher information}
%{\it Bounding mutual information by Fisher information.---}
The estimation of a parameter $\var$ from measurements is described
by two probabilities: the prior distribution of the parameter
$p(\var)$ and the conditional probability $p(x|\var)$ of the measurement
results $x$.  Together they constitute the joint probability
$p(x,\var)=p(x|\var)p(\var)$. The MI between random variables $\X$ and $\Var$ is
$I(\X,\Var):= H(\Var)-H(\Var|X)$, where
$H(\Var)=-\int d\var\:p(\var){\ln}p(\var)$ is the entropy of
$p(\var)$ and $H(\Var|\X)$ is the conditional entropy, given
$x$~\footnote{Note that, as we are dealing with continuous variables, the entropies may be negative.}. MI tells us the amount of information (in nats) about $\var$
obtained from $x$. The FI in given point $\var$ is
$F(\var)=\sum_x
\frac{1}{p(x|\var)}\left(\frac{dp(x|\var)}{d\var}\right)^2$ and tells
us how much information on $\var$ is contained in $p(x|\var)$. The
following two theorems relate MI to FI.

\begin{theorem}
\label{thm:1}
  If the parameter is guaranteed to lie inside of a finite-size set, namely $\t{supp}\,p(\var)\subseteq [a,b]$, then 
\begin{equation}
\label{eq:xilu}
 I(\X,\Var)\leq {\ln}\left(1+\frac{1}{2}\int_{a}^b d\var\sqrt{F(\var)}\right).
\end{equation}
\end{theorem}
Note especially, that there are no further constraints on $p(\var)$ -- it is an arbitrary probability density, continuous or discrete (or a combination of these two), as long as it takes value $0$ outside of $[a,b]$. This inequality holds for any $p(x|\var)$, so in the context of
quantum metrology, by replacing the FI with the quantum Fisher
information, one obtains a bound valid for any possible
measurement. Moreover the integral is invariant for reparametrization
of the probabilities $\var \to \var'$ (with appropriately modified limits of integration), as $F(\var')=F(\var)(\frac{d\var}{d\var'})^2$, so $\sqrt{F(\var')}d\var'=\sqrt{F(\var)}d\var$. In a broader context, this feature is crucial in defining non-informative prior distributions, the Jeffreys' prior~\cite{clarke1994jeffreys}. Yet another way to understand reparametrization-invariance of \eqref{eq:xilu} is to notice, that the FI is known to define a
metric corresponding to an Euclidean distance between
$\sqrt{p(x|\var)}$ and
$\sqrt{p(x|\var+d\var)}$~\cite{chen2022information}, so the above
integral is just the size of the region $[a,b]$ in this metric.

If the support of $p(\var)$ is not bounded, the following holds:

\begin{theorem}
\label{thm:2}
For arbitrary differentiable probability distribution $p(\var)$, we have
\begin{equation}
\label{eq:xiluprior}
	I(\X,\Var)\leq {\ln}\left(\frac{1}{2}\int_{-\infty}^{+\infty} d\var \sqrt{F(\var)p(\var)^2+\dot p(\var)^2}\right)+H(\Var),
\end{equation}
\end{theorem}
\noindent where $\dot p=dp/d\var$ and the $\ln$ part is obviously an
upper bound to $-H(\Var|x)$.  In contrast to the previous bound, this is not reparametrization invariant (indeed, it is well known
that the differential entropy $H(\Var)$ is defined modulo an arbitrary
additive constant). So, the upper bound, and also its tightness, will
depend on the parametrization.

\textit{Proof of the two bounds.} The mutual information may be written as:
\begin{multline}
I(\X,\Var)=\sum_x\int_{-\infty}^{+\infty} d\var\: p(x,\var)\ln\frac{p(x,\var)}{p(x)p(\var)}=\\
\sum_x\int\limits_{-\infty}^{+\infty} d\var\: p(x,\var)\ln \frac{p(x|\var)f(\var)}{p(x)}-\sum_x\int\limits_{-\infty}^{+\infty} d\var\: p(x,\var)\ln f(\var),
\end{multline}
where $f(\var)$ is an arbitrary non-negative function satisfying $\t{supp}\,f(\var)\supseteq \t{supp} \,p(\var)$. The first element of the above may be bounded by:
\begin{multline}
\label{eq:maxq}
%\sum_x\int d\var\:p(x,\var)\ln \frac{p(x|\var)f(\var)}{p(x)}\\
\leq \sum_x p(x)\ln \max_\var \frac{p(x|\var)f(\var)}{p(x)}
\\
\leq \max_{q_x:\sum_xq_x=1}\left(\sum_x q_x\ln \max_\var \frac{p(x|\var)f(\var)}{q_x}\right)\\
\leq \ln \sum_x \max_\var p(x|\var)f(\var),
\end{multline}
where the last inequality comes from  direct maximization over $q_x$, which is obtained for $q_x=\frac{p(x|\var)f(\var)}{\sum_x \max_\var p(x|\var)f(\var)}$ (see App. \ref{app:derivation}). If $f(\var)$ goes to $0$ for $\var\to\pm\infty$, we have:
\begin{equation}
\sum_x\max_\var p(x|\var)f(\var)\leq \frac{1}{2}\int_{-\infty}^{+\infty} d\var\sum_x \left|\frac{d(p(x|\var)f(\var))}{d\var}\right|.
\end{equation}
From Cauchy's inequality for vectors $\sqrt{p(x|\var)}$ and $\sqrt{\left|\frac{d(p(x|\var)f(\var))}{d\var}\right|^2\frac{1}{p(x|\var)}}$ (treated as a functions of single parameter $x$, with fixed $\var$) we have (see App. \ref{app:derivation}):
\begin{equation}
\label{eq:cauchy}
\sum_x \left|\frac{d(p(x|\var)f(\var))}{d\var}\right|\leq\\ \sqrt{F(\var)f(\var)^2+\dot f(\var)^2}.
\end{equation}
Combining all above one may get:
\begin{multline}
\label{eq:xiluf}
I(\X,\Var)\leq \ln\left(\frac{1}{2}\int_{-\infty}^{+\infty}d\var\sqrt{F(\var)f(\var)^2+\dot f(\var)^2}\right)+\\
-\int_{-\infty}^{+\infty} d\var\: p(\var)\ln f(\var),
\end{multline}
which may be further optimized over choice $f(\var)$ (going to zero for $\var\to\pm\infty)$. Note especially, that for $f(\var)=p(\var)$, we get \eqref{eq:xiluprior}. Alternatively, for $f(\var)=1$ on $[a,b]$ and $0$ outside (so $\dot f(\var)=\delta(\var-a)-\delta(\var-b)$), we obtain \eqref{eq:xilu}, see App. \ref{app:ab} for more details. $\square$

\sectiontext{From local to global estimation}
%{\it Bound for Bayesian cost.---}
The Fisher information, via the Cram{\'e}r-Rao inequality, constitutes a lower bound for the variance for any locally unbiased estimator and it is a meaningful tool in the situation when the number of repetitions of the estimation protocol is large. However, if the number of measurement repetitions is finite, properly including prior knowledge about the parameter may lead, when averaged, to a smaller variance (as the optimal estimator may not satisfy the local unbiasedness condition). It may be quantified by the general relation between averaged Bayesian cost $\overline{\Delta^2\tilde\var}:=\int d\var\:p(\var) \int d\tilde\var\:
p(\tilde\var|\var)(\tilde\var-\var)^2$, the entropy $H(\Var)$ of the prior distribution, and MI~\cite{cover2006elements,hall2012does}:%\footnote{Note a  typo in Eq. (11) in \cite{hall2012does} (missing square over  $\epsilon$)}
%$\frac{1}{2}{\ln}[2\pi e \overline{\Delta^2\tilde\var}]\geq H(\var|x)$
\begin{equation}
\label{eq:varentr}
\overline{\Delta^2\tilde\var}\geq \frac{e^{2H(\Var|X)}}{2\pi e}=\frac{e^{2H(\Var)}e^{-2I(\X,\Var)}}{2\pi e},
\end{equation}
Note that the above is tight iff all a posteriori distributions $p(\var|x)$ are Gaussian with the same variance. The above inequality in a clear manner separates the impact of the initial knowledge $-H(\Var)$ from the knowledge obtained from the experiment $I(\X,\Var)$. Note, that while $I(\X,\Var)$ is reparametrization invariant, both $\overline{\Delta^2\tilde\var}$ and $e^{2H(\Var)}$ behave under reparametrization in a similar way, making all inequality consistent. By applying the bound \eqref{eq:xilu} to the above we obtain 
\begin{equation}
  \overline{\Delta^2\tilde\var}\geq \frac{e^{2H(\Var)}}{2\pi e}\frac{1}{(1+\frac{1}{2}\int_a^b d\var\: \sqrt{F(\var)})^2}.
  \label{result}
\end{equation}
Note that the prior information affects not only $H(\Var)$, but also the range of the integral in the denominator. The better the parameter is known from the beginning, the more difficult it is to acquire significant new information from measurements. In
particular, for the relevant case of a rectangle prior of width $d$
with FI constant over the prior, one still obtains a significant
bound, contrary to the van Trees case (see next section for discussion):
 \begin{equation}
 \label{eq:bayesianflat}
\overline{\Delta^2\tilde\var}\geq \frac{2}{\pi e}\frac{1}{(2/d+\sqrt{F})^2}.
\end{equation}
As expected, in the limit of many repetitions $N$ of the estimation
procedure, the prior knowledge becomes irrelevant, since $F$ typically
grows linearly with $N$.
%(or quadratically, if the Heisenberg bound isattained).
A disadvantage of this bound is that it is not
asymptotically tight, because of the multiplicative factor $2/\pi e$.

Of course, a further bound for the Bayesian cost can be obtained also
for unbounded-support $p(\var)$ by inserting
\eqref{eq:xiluprior} into \eqref{eq:varentr}: 
\begin{equation}\label{result1}
\overline{\Delta^2\tilde\var}\geq
\frac{2}{\pi e}\frac{1}{(\int_{-\infty}^{+\infty} d\var
\sqrt{F(\var)p(\var)^2+\dot p(\var)^2})^2},
\end{equation}
which, for example, for Gaussian priors leads to a bound qualitatively similar to the van Trees one (see App. \ref{app:gaussian}). Note that for specific case of rectangle prior both bounds \eqref{result} and \eqref{result1} leads to the same results (see App. \ref{app:ab}).

Our results (\ref{result}) and (\ref{result1}) provide bounds to the
quadratic Bayesian cost. But a relation similar to \eqref{eq:varentr}
may be derived for any moment (see Lemma B.1. in
\cite{chen2024l_q}). Therefore, our bounds to MI imply a bound for the
Bayesian cost with an arbitrary cost function, assuming that it may be
expanded in Taylor series. These show that the bound for the MI is
more informative than any bound to Bayesian cost.

\sectiontext{Relation to the Efroimovich
and van Trees inequalities}
\label{sec:relation}
We will now compare our results with the bounds that exist in the literature. MI may be related to FI via Efroimovich relation \cite{efroimovich1980information} (see also
\cite{aras2019family} some generalizations and \cite[Corollary 3]{lee2022new} for an alternative proof):
\begin{equation}
\begin{split}
\label{eq:effinf}
I(\X,\Var)\leq& \frac{1}{2}\ln\left[\frac{1}{2\pi e}\left(\int_{-\infty}^{+\infty} d\var F(\var)p(\var)+P\right)\right]+H(\Var),\\
&{\rm with}\quad P=\int_{-\infty}^{+\infty} d\var\frac{1}{p(\var)}\left(\frac{dp(\var)}{d\var}\right)^2,
\end{split}
\end{equation}
where $P$ can be interpreted as the information included in the
prior. 
While for Gaussian prior the quantity $P$ is a reasonable measure of
information (the inverse of the variance), it is completely
unreasonable for the rectangle distribution, where $P$ diverges. In general, $P$ depends more on the sharpness of the prior distribution on its
edges than on its actual width, which is more an artifact of the
derivation of the bound, rather than a well-motivated
feature. Overcoming this problem was also discussed
in~\cite{aras2019family,lee2022new}.

Introducing Efroimovich’s
inequality into \eqref{eq:varentr} we find the van
Trees one~\cite{gill1995}\footnote{Note, that van Trees' inequality has been derived first~\cite{schutzenberger1957generalization} and it has been noticed to be a consequence of Efroimovich’s inequality only recently.}:
\begin{equation}
\label{eq:vantrees}
\overline{\Delta^2\tilde\var}\geq \frac{1}{\int_{-\infty}^{+\infty} d\var\:F(\var)p(\var)+P},
\end{equation}
which again suffers sharp $p(\var)$. See also~\cite{bobrovsky1987some,gill1995,tsang2020physics} for more advanced versions of the inequality that solve this problem.

This issue does not appear in our bounds (\ref{eq:xilu}) and
(\ref{eq:xiluprior}). Indeed, the
impact of $\dot p(\var)$ can be bounded by
$\frac{1}{2}\int_{-\infty}^{+\infty} d\var |\dot p(\var)|$, as discussed in App. \ref{app:ab}. In particular, for any
prior concentrated around one region (more formally: the prior where
the derivative $\dot p(\var)$ changes its sign only once), it may be
bounded by constant $\max_\var p(\var)$, no matter how sharply the prior
changes.

At last, let us discuss the asymptotical behavior of the bounds for MI. We start with showing asymptotic saturability of Efroimovich's inequality (\ref{eq:effinf}) in the case when newly obtained data dominate initial knowledge.
Assuming that $P$ does not diverge, and that the probability
$p(x,\var)$ is not degenerate
(i.e. $\forall_{\var'\neq\var}\exists_xp(x,\var')\not= p(x,\var)$), then (\ref{eq:effinf}) is tight asymptotically in the
number of repetitions $N$ of the estimation, and saturable using the
maximum likelihood estimator. Indeed, first, the FI increases linearly
with $N$, so asymptotically the impact of $P$ in the upper bound
becomes negligible. Second, from the asymptotic normality of the
maximum likelihood estimator, the difference between $\tilde \var_{\t ML}$ and the true value of the parameter $\var$ converge to normal distribution of variance $1/F(\var)$, so we have~\cite{brunel1998mutual}
\begin{equation}\label{ef1}
H(\Var|\tilde \Var_{\t ML})\to -\frac{1}{2}\ln \left[\frac{1}{2\pi e} \int_{-\infty}^{+\infty} d\var F(\var)p(\var)\right].
\end{equation}
Third, assigning the estimator's value to the measurement result
$\tilde \var_{\t ML}(x)$ may only decrease the MI, namely
$I(\tilde \Var_{\t ML},\Var)\leq I(\X,\Var)$. This lower bound, through
\eqref{ef1} converges to the upper bound \eqref{eq:effinf} in the limit, where $P$ is negligible.

From this, we see that for regular priors (finite $P$) the bound \eqref{eq:xiluprior} is not asymptotically tight, because of the multiplicative factors inside the $\ln$.

%\begin{figure}[t!]
%\includegraphics[width=0.45\textwidth]{schemes_noisy.pdf}
%\caption{Different possible quantum estimation schemes: $N$ elementary   gates (black boxes) are used on separably prepared probes in SEP; in  a sequential manner in SEQ; starting from an entangled input state  (shaded box) in EN; in an adaptive way in AD ($V_i$ representing joint unitaries). Each subsequent one
  %$is   more powerful than/or   includes all previous ones as special  cases. The gray cups on the right represent a (possibly joint)  measurement.}\label{fig:all}\end{figure}

\sectiontext{The bound for Holevo information} Now we consider the quantum case, where the conditional probability distribution of the measurement output is given via
the Born rule $p(x|\var)=\tr(\rho_\var M_x)$, where $\rho_\var$ is density matrix of the quantum state, $M_x$ is a POVM element ($M_x\geq 0$,
$\sum_x M_x=\openone$). Further, we assume that the state $\rho_\var$ is drawn with probability distribution $p(\var)$. The mutual information $I(\X,\Var)$, maximized over the choice of the measurement $\{M_x\}_x$, is bounded by Holevo information~\cite{wilde2011classical}. The following quantum version of \autoref{thm:1} and \autoref{thm:2} holds.

\begin{theorem}\label{thm:qmi_bound}
Given a family of quantum states $\rho_\var$ (differentiable with respect to $\var$), appearing with probability distribution $p(\var)$ with $\t{supp}\,p(\var)\subseteq [a,b]$, the Holevo information $\chi$ is bounded by:
\begin{equation}
\label{eq:quantum1}
    \chi\leq \ln\left(1+\tfrac{1}{2}\int_{a}^b d\var \sqrt{F_Q[\rho_\var]}\right),
\end{equation}
with $F_Q$ the quantum Fisher information of $\rho_\var$ and $\chi := S\left(
            \int p(\var) \rho_{\var} d\var
        \right)
        -
        \int p(\var) S(\rho_{\var}) d\var$, with $S(\rho):=-\tr(\rho\log\rho)$ the von Neuman entropy.

For any differentiable probability distribution $p(\var)$, the Holevo information is bounded by:
\begin{equation}
\label{eq:quantum2}
	\chi\leq {\ln}\left(\frac{1}{2}\int_{-\infty}^{+\infty} d\var \sqrt{F_Q[\rho_\var]p(\var)^2+\dot p(\var)^2}\right)+H(\Var).
\end{equation}
\end{theorem}
\textit{Proof.} See App. \ref{app:qmi_bound} $\square$.

\sectiontext{Applications to quantum metrology}
%{\it Applying bound to quantum metrology.---}
To show the usefulness of the newly derived bounds, we now show how they can be used to give bounds to noisy quantum metrology for mutual information.

Consider the $\var$-dependent CPTP map $\Lambda_\var[\cdot]$ that
noisily encodes the parameter $\var$ onto a quantum probe. The phase estimation problem is a typical
example. Consider the simple case of qubit probes, where the parameter
$\var$ is encoded through a unitary
$U_{\var}=\ket{0}\bra{0}+e^{i\var}\ket{1}\bra{1}$, followed by
some kind of noise (e.g.~dephasing, erasing) which together constitute
the map $\Lambda_\var[\cdot]$, which we will further treat as a quantum gate. The question is: what maximal MI may be obtained with the usage of $N$ gates in an optimal way?

%\wg{Simulanously, even if uniform prior is often seen as "no information", the fact, that the set of possible values of the parameter has a finite size $2\pi$ has a significant impact on the minimal obtainable variance (especially in situation, where newly obtained data is small).}

In~\cite{hassani2017digital} the Heisenberg Scaling (HS) for MI has
been defined as $I\propto{\ln} N$ and the standard quantum limit as
$I\propto\frac{1}{2}{\ln} N$ (in analogy to the scaling of RMSE
$\Delta\var\propto1/N$ and $\Delta\var\propto1/\sqrt{N}$
respectively). In the case of noiseless estimation, it is known that using gates separately allows at most standard quantum limits, while more complicated schemes, including entanglement or multipassing, allow for obtaining HS (for example via QPEA algorithm~\cite{hassani2017digital}). However, the
noisy case has not been discussed up to now while using MI as a figure
of merit.

Note that to obtain HS it is necessary to use all available resources jointly in a single experiment realization, therefore the argument based on the asymptotic efficiency of the maximum likelihood estimator cannot be applied here. Therefore, even if intuitively we expect that the inability to obtain HS in FI should imply an inability to obtain HS also for MI, it is not trivial to prove it formally in a general case. 
Note also, that in this model
$\var\in[0,2\pi[$, and the Efroimovich’s inequality or van Trees inequality cannot be applied
to a uniform prior.

We can now apply the bounds derived above, e.g.~\eqref{eq:xilu}, to
transfer the highly nontrivial theory of noisy-channel estimation
theory from the FI to the MI formalism. For FI, the general necessary
and sufficient conditions for the obtainability of HS in the
estimation of a given channel are known, expressed in terms of Kraus
operators~\cite{demkowicz2014using,kurdzialek2022using,zhou2021asymptotic}
or Lindblad operators~\cite{demkowicz2017adaptive,layden2018ancilla}.
Moreover, the resulting bounds are known to be saturable in the
asymptotic limit $N\to\infty$ of many
gates~\cite{zhou2021asymptotic}. For example, for a mentioned model of
phase estimation, it was shown that in the presence of dephasing or
erasure noise, the FI obtainable by any protocol involving $N$ gates
is bounded from above by $F\leq N \eta/(1-\eta)$
\cite{demkowicz2014using}%\footnote{Note that the factor $(2\pi)^2$  comes from using a different parametrization in this paper)}
, where
$\eta$ is the noise parameter ($\eta\to 1$ in the noiseless case,
$\eta\to 0$ for maximum noise). The bound remains valid even for the most general adaptive scheme, including entanglement with arbitrary large ancilla and applying unitary control after each action of gate. Applying this bound to \eqref{eq:xilu}
we immediately find
$I(\X,\Var)\leq {\ln}(1+\pi\sqrt{N}\sqrt{\eta/(1-\eta)})$ and using \eqref{eq:bayesianflat} we have 
$
\overline{\Delta^2\tilde\var}\geq 2/(\pi e)(1/\pi+\sqrt{N}\sqrt{\eta(1-\eta)})^{-2}.
$

Even if the bound is not tight
(there is an additive constant gap), it implies a standard quantum
limit asymptotic scaling for all $\eta<1$, and also gives the scaling
with the intensity of the noise (see App. \ref{app:noisy} for more
details). Similar results can be obtained easily using different
bounds to
FI~\cite{escher2011general,demkowicz2012elusive,knysh2014true}. Naturally,
this does not rule out quantum enhancements: for sufficiently large
$\eta$ one might still find an arbitrarily large {\em constant} gain
over the SQL.

These will provide only the upper bounds to MI and so
they can only be used to exclude the possibility of attaining the
HS, but never to prove its achievability. While there
exist lower bounds on the MI expressed in terms of the FI in the
literature, they all require additional assumptions and a general
discussion is impossible. 
In the noiseless case, it may be exemplified by the N00N state which
allows the HS for the FI, but not for the MI, due to its periodicity with period $1/N$ (indeed, irrespectively of the value of $\var$, the N00N state
belongs to a two-dimensional subspace of the full Hilbert space, so it can carry at most one bit of information
$I\leqslant{\ln}(2)$). This demonstrates that, while the N00N state is highly effective for local estimation, it presents significant challenges for global estimation and requires careful consideration in its application~\cite{higgins2009demonstrating}.

\sectiontext{Applications to communication}
Beyond the problems of estimation and metrology, our bound is also useful for transferring the theorems derived in the metrology context to communication problems. Consider a situation where Alice employs the parameter $\var\in[a,b]$ as an alphabet for communicating with a state $\rho_\var$. Alice has the freedom to choose if she will use all values of $\var$ or restrict herself to a discrete subset. From \eqref{eq:xilu} the MI between Alice and Bob is then bounded by
$I(A,B)\leq \ln\left(1+\tfrac{1}{2}\int_{a}^b d\var \sqrt{F_Q[\rho_\var]}\right)$. The bound works independently of what messages (and with what probabilities) Alice is sending. Moreover, one can use \eqref{eq:xilu} to give bounds to the classical capacity of a quantum or classical channel when the decoding strategy is fixed and its FI is known.

\sectiontext{Conclusions}
%{\it Conclusions.---}
In conclusion, we have derived two bounds, Eqs.~(\ref{eq:xilu}) and
(\ref{eq:xiluprior}) for the MI in terms of the FI. We used these
bounds to give two extensions of the van Trees inequality,
Eqs.~(\ref{result}) and (\ref{result1}). We discussed the relation of
all these bounds to the Efroimovich inequality (\ref{eq:effinf}). We also derived the quantum version of these bounds, Eqs.~(\ref{eq:quantum1}) and~(\ref{eq:quantum2}), where the Holevo information is bounded in terms of QFI. To prove the usefulness of
these bounds, we have shown that they can be used to extend to the
MI-based quantum metrology many highly nontrivial results known for
the FI case. They also constitute the bounds on communication in terms of FI. See also \cite{lu2024number} for alternative bounds linking MI with FI.

\sectiontext{Acknowledgements} We thank Giovanni Chesi, Denis Vasilyev, Pavel Sekatski, and Jacek Krajczok for the fruitful
discussions.  %L.M.
C.M., W.G.~acknowledge financial support from the
   U.S. DoE, National Quantum Information Science Research Centers,
   Superconducting Quantum Materials and Systems Center (SQMS) under
   contract number DE-AC02-07CH11359, X.L. acknowledges the National Natural Science Foundation of China under Grant Nos.62272406, 61932018 and Zhejiang University for funding, and the University of Pavia for
   hospitality, C.M. acknowledges support from the EU H2020 QuantERA project QuICHE and from the PNRR MUR Project PE0000023-NQSTI. L.M. acknowledges support from the
   %PNRR MUR Project CN0000013-ICSC, and from 
   PRIN2022 CUP 2022RATBS4.

\bibliography{biblio}

\appendix

\section{Derivation of inequalities (\ref{eq:maxq}) and (\ref{eq:cauchy})}
\label{app:derivation}

First we derive inequality (\ref{eq:maxq}), i.e. we perform direct maximization of
\begin{equation}
\max_{q_x:\sum_xq_x=1}\left(\sum_x q_x\ln \max_\var \frac{p(x|\var)f(\var)}{q_x}\right)
\end{equation}
by using the Lagrange multiplier method. The condition for local extremum with constrain $\sum q_x=1$ is:
\begin{multline}
\frac{\partial}{\partial q_x}\left[\sum_x q_x\ln \max_\var \frac{p(x|\var)f(\var)}{q_x}-\lambda\left(\Big(\sum_x q_x\big)-1\right)\right]=\\\ln \max_\var \frac{p(x|\var)f(\var)}{q_x}-1-\lambda=0,
\end{multline}
which imposes that $\ln \max_\var \frac{p(x|\var)f(\var)}{q_x}$ are the same for all $x$. Together with $\sum q_x=1$ is gives $q_x=\frac{p(x|\var)f(\var)}{\sum_x \max_\var p(x|\var)f(\var)}$. Since it is the only stationary point and the function takes smaller values outside of this point, this is a global maximum, which was to be proven.

Next, we derive (\ref{eq:cauchy}), using Cauchy's inequality for vectors $\sqrt{\left|\frac{d(p(x|\var)f(\var))}{d\var}\right|^2\frac{1}{p(x|\var)}}$ and $\sqrt{p(x|\var)}$.
\begin{equation}
\label{eq:cauchyder}
\begin{aligned}
	&\sum_x \left|\frac{d(p(x|\var)f(\var))}{d\var}\right|=\\
 &=\sum_x \left[\sqrt{\left|\frac{d(p(x|\var)f(\var))}{d\var}\right|^2\frac{1}{p(x|\var)}}\sqrt{p(x|\var)}\right]\\
 &\leq \sqrt{\sum_x\left|\frac{d(p(x|\var)f(\var))}{d\var}\right|^2\frac{1}{p(x|\var)}}\sqrt{\sum_xp(x|\var)}\\
 &=\sqrt{\sum_x\left|\frac{d(p(x|\var)f(\var))}{d\var}\right|^2\frac{1}{p(x|\var)}}\\
 &\sqrt{\sum_x \left[\frac{\dot p(x|\var)^2}{p(x|\var)}f(\var)^2+2\dot p(x|\var)f(\var)\dot f(\var)+p(x|\var)\dot f(\var)^2\right]}\\
 &=\sqrt{F(\var)f(\var)^2+\dot f(\var)^2}.
\end{aligned}
\end{equation}
where we used $\sum_x p(x|\var) = 1$ and $\sum_x \dot{p}(x|\var) = 0$.

\section{Obtaining \eqref{eq:xilu}}
\label{app:ab}
Here we show how to obtain \eqref{eq:xilu} from \eqref{eq:xiluf}:
\begin{multline}
I(\X,\Var)\leq \ln\left(\frac{1}{2}\int_{-\infty}^{+\infty}d\var\sqrt{F(\var)f(\var)^2+\dot f(\var)^2}\right)+\\
-\int_{-\infty}^{+\infty} d\var\: p(\var)\ln f(\var).
\end{multline}
First, note that in general $\sqrt{x+y}\leq \sqrt{x}+\sqrt{y}$, so the above expression implies:
\begin{multline}
\label{eq:step}
I(\X,\Var)\leq \ln\Big(\frac{1}{2}\int_{-\infty}^{+\infty}d\var\sqrt{F(\var)f(\var)^2}
+\frac{1}{2}\int_{-\infty}^{+\infty}d\var|\dot f(\var)|\Big)\\
-\int_{-\infty}^{+\infty} d\var\: p(\var)\ln f(\var).
\end{multline}
Next, choosing $f(\var)=1$ on $[a,b]$ and $0$ outside (so $\dot f(\var)=\delta(\var-a)-\delta(\var-b)$), we have:
\begin{equation}
\label{eq:final}
I(\X,\Var)\leq \ln \left(1+\frac{1}{2}\int_{a}^{b}d\var\sqrt{F(\var)}
\right).
\end{equation}

Suppose one would prefer a more rigorous approach (avoiding taking the derivative of the rectangle function). In that case, one may always consider a family of functions $f_\epsilon(\var)$ satisfying: $f_\epsilon(\var)=1$ on $[a,b]$, $f_\epsilon(\var)=0$ outside of $[a-\epsilon,b+\epsilon]$ and $f_\epsilon(\var)$ is monotonic and differentiable on both $[a-\epsilon,a]$ and $[b,b+\epsilon]$. Then, assuming that $F(\var)$ does not diverge for any $\var$, in the limit $\epsilon\to 0$, \eqref{eq:step} converge to \eqref{eq:final}.

Note that \eqref{eq:step} implies also, that an impact of $\dot p(\var)$ in \eqref{eq:xiluprior} may be bound by $\frac{1}{2}\int_{-\infty}^{\infty}|\dot p(\var)|$. Especially, for a specific case of rectangle prior ($p(\var)=1/(b-a)$ on $[a,b]$ and $0$ outside), \eqref{eq:xiluprior} therefore leads to \eqref{eq:xilu}, as:
\begin{multline}
	I(\X,\Var)\leq {\ln}\left(\frac{1}{2}\int_{-\infty}^{+\infty} d\var \sqrt{F(\var)p(\var)^2+\dot p(\var)^2}\right)+H(\Var)\\
 \leq{\ln}\left(\frac{1}{b-a}+\frac{1}{2(b-a)}\int_{a}^{b} d\var \sqrt{F(\var)}\right)+\ln(b-a)\\
 ={\ln}\left(1+\frac{1}{2}\int_{a}^{b} d\var \sqrt{F(\var)}\right).
\end{multline}

\section{Applying the bound for Gaussian distribution}
\label{app:gaussian}
Consider an a priori distribution $p(\var)=\frac{1}{\sqrt{2\pi\sigma^2}}\exp(-\frac{\var^2}{2\sigma^2})$. Substituting \eqref{eq:xiluprior} into \eqref{eq:varentr} we obtain:
\begin{equation}
\overline{\Delta^2\tilde\var}\geq \frac{2}{\pi e}\frac{1}{(\int d\var\sqrt{F(\var)+\var^2/\sigma^4}p(\var))^2}.
\end{equation}
Further, assuming that $F(\var)$ does not change with $\var$, the integral may be calculated analiticaly as $\frac{\sqrt{2}}{\sigma} U[-\frac{1}{2}, 0, \frac{F\sigma^2}{2}]$, where $U(\cdot)$ is Tricomi confluent hypergeometric function.

For a slightly less tight, but simpler bound
for the case where FI does not depend on $\var$, note that after introducing the variable $\eta=\var^2$, the quantity $\sqrt{F+\eta/\sigma^4}$ is concave as a function of $\eta$, so $\mathbb E[\sqrt{F+\eta/\sigma^4}]\leq \sqrt{F+\mathbb E[\eta]/\sigma^4}$ (with $\mathbb E[\cdot]$ denoting averaging), so the integral may be bounded from above by $\sqrt{F+1/\sigma^2}$ (numerical results also show that this is a good approximation for the exact result for $F\sigma^2\gg 1$). For any $F,\sigma$ we have:
\begin{equation}
\overline{\Delta^2\tilde\var}\geq \frac{2}{\pi e}\frac{1}{F+1/\sigma^2},
\end{equation}
while from the van  Trees inequality we have \eqref{eq:vantrees} (taken for constant $F$):
\begin{equation}
\overline{\Delta^2\tilde\var}\geq \frac{1}{F+1/\sigma^2}.
\end{equation}
Then, for this specific case, our bound is less tight, because of the $2/\pi e$ multiplicative factor.

\section{Mutual information in occurrence of noise}
\label{app:noisy}

\begin{figure}
    \centering
    \includegraphics[width=0.45\textwidth]{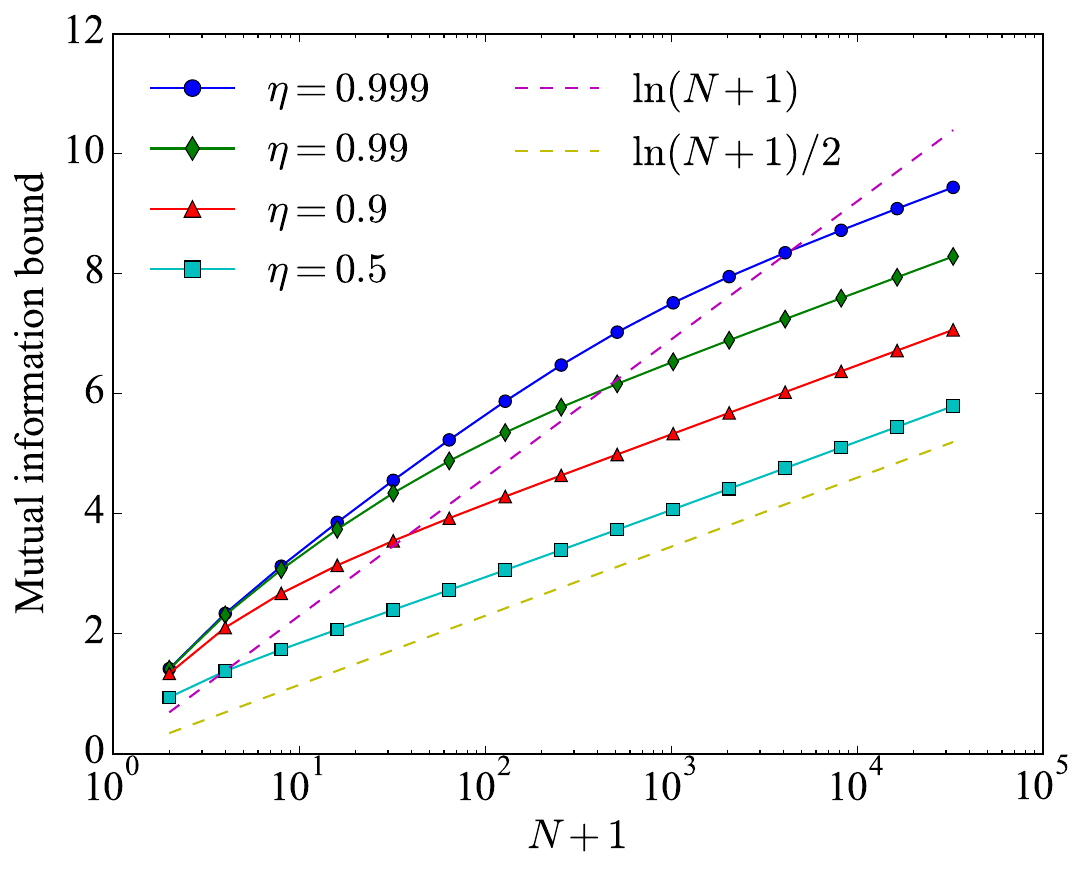}
    \caption{The mutual information bound given by \eqref{eq:xilu} with maximum Fisher information, for the dephasing channel and amplitude damping channel. The two dashed lines help to compare the bound with HS and SQL.}
    \label{fig:fisher_bound_noise}
\end{figure}

We now give a global bound to the mutual information in terms of $N$, the number of calls to the phase gate $U_\var=\ket{0}\bra{0}+e^{i \var}\ket{1}\bra{1}$, in the presence of dephasing and amplitude damping noise.

In our noise model, each unitary gate $U_\var$ is replaced by the noisy gate acting on the density operator as
\begin{equation}
    \Lambda_\var: \rho \mapsto \sum_k K_k U_\var \rho U_\var^\dagger K_k^{\dagger},
\end{equation}
with $K_k$ being the Kraus operators.

For dephasing noise the Kraus operators are
\begin{equation}
    K_0 = \openone \left(
        \frac{1+\sqrt{\eta}}{2}
    \right)^{1/2},
    \qquad
    K_1 = \sigma_z \left(
        \frac{1-\sqrt{\eta}}{2}
    \right)^{1/2},
\end{equation}
where $\openone=\ket{0}\bra{0}+\ket{1}\bra{1}$, $\sigma_z=\ket{0}\bra{0}-\ket{1}\bra{1}$, and $\eta$ is the noise parameter.
Also, the Kraus operators in the presence of amplitude damping noise are
\begin{equation}
    K_0 = \begin{pmatrix}
        1 & 0 \\
        0 & \sqrt{\eta}
    \end{pmatrix},
    \quad
    K_1 = \begin{pmatrix}
        0 & \sqrt{1-\eta} \\
        0 & 0
    \end{pmatrix}.
\end{equation}

Finally, for erasure noise the Kraus operators are
\begin{equation}
\begin{gathered}
    K_0 = \begin{pmatrix}
        \sqrt{\eta} & 0 & 0 \\
        0 & \sqrt{\eta} & 0 \\
        0 & 0 & 0
    \end{pmatrix},
    \quad
    K_1 = \begin{pmatrix}
        0 & 0 & 0 \\
        0 & 0 & 0 \\
        0 & 0 & 1
    \end{pmatrix},
    \\
    K_2 = \begin{pmatrix}
        0 & 0 & 0 \\
        0 & 0 & 0 \\
        \sqrt{1-\eta} & 0 & 0
    \end{pmatrix},
    \quad
    K_3 = \begin{pmatrix}
        0 & 0 & 0 \\
        0 & 0 & 0 \\
        0 & \sqrt{1-\eta} & 0
    \end{pmatrix},
\end{gathered}
\end{equation}
where a third dimension is added to indicate loss of phase information.

In this notation, an asymptotic upper bound to the Fisher information is~\cite{demkowicz2014using,kurdzialek2022using},
\begin{equation}
    F \leq NF_{\mathrm{as}},
\end{equation}
where $F_{\mathrm{as}}=\eta/(1-\eta)$ for both dephasing and erasure noise with the most general AD strategy, as well as amplitude damping noise with EN strategy~\cite{knysh2014true}.
Combining the Fisher information bound with \eqref{eq:xilu}, we get a global upper bound to the mutual information,
\begin{equation}
    I(\X,\Var) \leq \ln\left(
        1 + \pi\sqrt{\frac{N\eta}{1-\eta}}
    \right).
    \label{eq:fisher_noise_asympt}
\end{equation}

For amplitude damping noise, the Fisher information bound is slightly different, and the mutual information bound can be derived similarly.

Especially, when limited to the EN strategy, a tighter bound to Fisher information for finite-$N$ is given by~\cite{kolodynski2013efficient},
\begin{equation}
    F \leq \frac{N F_{\mathrm{as}}}{1+\frac{F_{\mathrm{as}}}{N}}.
\end{equation}

Combining the Fisher information bound with \eqref{eq:xilu}, we get a global upper bound to the mutual information
\begin{equation}
    I(\X,\Var) \leq \ln\left(
        1 + \pi\sqrt{\frac{N \eta/(1-\eta)}{1+\eta/N(1-\eta)}}
    \right),
    \label{eq:fisher_bound_noise}
\end{equation}
for all these three noise models.
We plot the right hand side of \eqref{eq:fisher_bound_noise} in \autoref{fig:fisher_bound_noise}, in which we can see a transition from the HS to the SQL as $N$ increases. The stronger the noise is, the earlier the transition happens.

\section{Holevo information bounded by Quantum Fisher Information -- proof of \autoref{thm:qmi_bound}}
\label{app:qmi_bound}

In this section, we prove \autoref{thm:qmi_bound}, using a similar way of reasoning as in the proof of \autoref{thm:1}, \autoref{thm:2}, adapted to the quantum version. Namely, we introduce function $f(\var)$ and prove stronger statement: 

\begin{theorem}
\label{thm:qmi_bound_f}
Given a family of quantum states $\rho_\var$ (differentiable with respect to $\var$), appearing with probability distribution $p(\var)$. For any non-negative function $f(\var)$ satisfying $\t{supp}\,f(\var)\supseteq \t{supp} \,p(\var)$ and vanishing at $\pm\infty$,
the Holevo information $\chi$ is bounded by:
    \begin{equation}\label{eq:qmi_bound}
    \begin{aligned}
        \chi \leq &
        \ln\left(\frac{1}{2}\intinf \dd\var\sqrt{F_Q[\rho_{\var}] f(\var)^2+\dot f(\var)^2}\right)
        \\ & - \intinf \dd\var\: p(\var)\ln f(\var),
    \end{aligned}
    \end{equation}
   with $F_Q$ the quantum Fisher information of $\rho_\var$ and $\chi := S\left(
            \int p(\var) \rho_{\var} d\var
        \right)
        -
        \int p(\var) S(\rho_{\var}) d\var$, with $S(\rho):=-\tr(\rho\log\rho)$ the von Neuman entropy.  
\end{theorem}

Then \autoref{thm:qmi_bound} comes as a collolary of \eqref{eq:qmi_bound} with $f(\var)=1/(b-a)$ on $[a,b]$ and $0$ outside or with $f(\var)=p(\var)$.

Before the proof we give the following lemmas.

\begin{lemma}\label{lem:tr_abs}
    For any two vectors $\ket{a},\ket{b}$,
    \begin{equation}
        \tr\big|
            \ketbra{a}{b} + \ketbra{b}{a}
        \big|
        \leq
        2\sqrt{
            \ip{a}{a} \ip{b}{b}
        },
    \label{eq:tr_abs}
    \end{equation}
    where $|A|:=\sqrt{A^\dagger A}$ for any operator $A$.
\end{lemma}

\textit{Proof of \autoref{lem:tr_abs}.}
The column space of the operator $\ketbra{a}{b} + \ketbra{b}{a}$ is $\mathcal{A}:=\mathrm{span}(\ket{a},\ket{b})$ of dimension at most 2.
When $\dim\mathcal{A}\leq 1$ the LHS becomes $2|\mathrm{Re}\ip{a}{b}|$, making it a trivial result.

When $\dim\mathcal{A}=2$, suppose $x\ket{a}+y\ket{b}\in\mathcal{A}$ is an eigenvector of $\ketbra{a}{b} + \ketbra{b}{a}$ with eigenvalue $\lambda$, then
\begin{equation}
    \lambda = \ip{b}{a} + \frac{y}{x} \ip{b}{b} = \ip{a}{b} + \frac{x}{y} \ip{a}{a}.
\end{equation}

By eliminating $y/x$ we get
\begin{equation}
    \lambda^2 - \big(
        \ip{a}{b}+\ip{b}{a}
    \big) \lambda - \big(
        \ip{a}{a}\ip{b}{b}-\ip{a}{b}\ip{b}{a}
    \big) = 0.
\end{equation}

Note that $\ip{a}{b}+\ip{b}{a}\in\mathbb{R}$ and $-\ip{a}{a}\ip{b}{b}+\ip{a}{b}\ip{b}{a}\le 0$, thus the equation has two real roots $\lambda_1,\lambda_2$ with opposite signs.
Moreover,
\begin{equation}
\begin{aligned}
    &
    \tr\big|
        \ketbra{a}{b} + \ketbra{b}{a}
    \big|
    \\ = &
    |\lambda_1| + |\lambda_2|
    \\ = &
    |\lambda_1 - \lambda_2|
    \\ = &
    \sqrt{(\lambda_1+\lambda_2)^2-4\lambda_1\lambda_2}
    \\ = &
    \sqrt{(\ip{a}{b}+\ip{b}{a})^2+4(\ip{a}{a}\ip{b}{b}-\ip{a}{b}\ip{b}{a})}
    \\ = &
    \sqrt{(\ip{a}{b}-\ip{b}{a})^2+4\ip{a}{a}\ip{b}{b}}
    \\ \leq &
    2\sqrt{\ip{a}{a}\ip{b}{b}},
\end{aligned}
\end{equation}
where in the last inequality we use $\ip{a}{b}-\ip{b}{a}\in i\mathbb{R}$.
$\square$

\begin{lemma}
    \label{lem:log_positive}
    Given positive operators $A,B$ satisfying $0<A\leq B$, i.e., $A$ is positive definite and $(B-A)$ is positive semi-definite, then
    \begin{equation}
        \ln(A) \leq \ln(B).
        \label{eq:log_positive}
    \end{equation}
\end{lemma}

\textit{Proof of \autoref{lem:log_positive}.}
The condition $0<A\leq B$ implies that $0<A^{1/2}\leq B^{1/2}$~\cite{lax07}.
By iterating it we know $0<A^{1/2^m}\leq B^{1/2^m}$ holds for any positive integer $m$.
Finally, by the identity $\ln(a)=\lim_{x\to 0}(a^x-1)/x$,
\begin{equation}
    \ln(A)=\lim_{m\to+\infty} 2^m (A^{1/2^m}-I).
\end{equation}
then \eqref{eq:log_positive} is obtained.
$\square$

\textit{Proof of \autoref{thm:qmi_bound_f}.}
We first consider the pure state case where $\rho_{\var}=\dyad{\psi_\var}$.
Write,
\begin{equation}
    \rho = \intinf p(\var) \rho_{\var} \dd\var.
\end{equation}
Then $\chi=S(\rho)$, and the quantum Fisher information is also simplified to,
\begin{equation}
    F_Q[\rho_{\var}] = 4\left[
        \ip{\dot\psi_{\var}}{\dot\psi_{\var}} - \ip{\dot\psi_{\var}}{\psi_{\var}}\ip{\psi_{\var}}{\dot\psi_{\var}}
    \right].
    \label{eq:qfi_original}
\end{equation}
Moreover, by adjusting the global phase of $\ket{\psi_{\var}}$ we can force $\ip{\psi_{\var}}{\dot\psi_{\var}}=0$ and simplify \eqref{eq:qfi_original} to,
\begin{equation}
    F_Q[\rho_{\var}] = 4\ip{\dot\psi_{\var}}{\dot\psi_{\var}}.
    \label{eq:qfi_simp}
\end{equation}

To prove \eqref{eq:qmi_bound}, we first shift the last term to the left side. For pure state (using $\chi=S(\rho)$), we therefore have:
\begin{equation}
\label{eq:firststep}
\begin{aligned}
    &\chi + \intinf \dd\var\: p(\var) \ln f(\var)
    \\=   &
    S(\rho) + \intinf \dd\var\: p(\var) \ln f(\var)
    \\ = &
    S(\rho) + \tr \intinf \dd\var\: p(\var)\rho_{\var} \ln [f(\var)\rho_{\var}].
\end{aligned}
\end{equation}
Now, to bound it from above we will find the matrix bounding $f(\var)\rho_{\var}$ from above (in the sense of matrix inequality).

To do that, we define an unnormalized vector
$\ket{\Psi_{\var}} = \sqrt{f(\var)} \ket{\psi_{\var}}$ and the operator:
\begin{equation}
    \trho = \frac{1}{2} \intinf \left|\ddvar\dyad{\Psi_{\var}}\right| \dd\var.
\end{equation}
For any $\var_0\in\mathbb{R}$,
\begin{equation}
\begin{aligned}
    &
    \trho
    \\ = &
    \frac{1}{2} \int_{-\infty}^{\var_0} \left|\ddvar\dyad{\Psi_{\var}}\right| \dd\var
    +
    \frac{1}{2} \int_{\var_0}^{+\infty} \left|\ddvar\dyad{\Psi_{\var}}\right| \dd\var
    \\ \ge &
    \frac{1}{2} \int_{-\infty}^{\var_0} \ddvar\dyad{\Psi_{\var}} \dd\var
    -
    \frac{1}{2} \int_{\var_0}^{+\infty} \ddvar\dyad{\Psi_{\var}} \dd\var
    \\ = &
    \dyad{\Psi_{\var}}
    =
    f(\var_0) \rho_{\var_0},
\end{aligned}
\end{equation}
where we use $f(\var)|_{\pm\infty}=0$.

By \autoref{lem:log_positive}, $\ln[f(\var)\rho_{\var}] \le \ln\trho$ in the support space of $\rho_{\var}$, thus we can further bound \eqref{eq:firststep} by:
\begin{equation}
\label{eq:secondstep}
\begin{aligned}
    &S(\rho) + \tr \intinf \dd\var\: p(\var)\rho_{\var} \ln [f(\var)\rho_{\var}]
    \\ \le &
    S(\rho) + \tr \intinf \dd\var\: p(\var)\rho_{\var} \ln \trho
    \\ = &
    S(\rho) + \tr (\rho \ln \trho)
    \\ = &
    -S(\rho||\trho^0) + \ln \tr \trho
    \\ \le &
    \ln \tr \trho.
\end{aligned}
\end{equation}
where $\trho^0:=\trho/\tr\trho$ is the normalized $\trho$ and $S(\rho||\trho^0):=\tr[\rho(\ln\rho-\ln\trho^0)]$ is the quantum relative entropy, which is always non-negative. While calculating $\tr \trho$ we can go with the trace under integral:
\begin{equation}
    \tr\trho
    =
    \frac{1}{2} \intinf \tr\left|\ddvar\dyad{\Psi_{\var}}\right| \dd\var,
\end{equation}
so RHS of \eqref{eq:secondstep} may be further bounded using \autoref{lem:tr_abs},
\begin{equation}
\begin{aligned}
    &\tr \left|\ddvar\dyad{\Psi_{\var}}\right|
    \\ = &
    \tr\big|\ketbra{\Psi_{\var}}{\dot\Psi_{\var}}+\ketbra{\dot\Psi_{\var}}{\Psi_{\var}}\big|
    \\ \le &
    2\sqrt{\ip{\Psi_{\var}}{\Psi_{\var}} \ip{\dot\Psi_{\var}}{\dot\Psi_{\var}}}
    \\ = &
    % \sqrt{f(\var)^2\ip{\dot\psi_{\var}}{\dot\psi_{\var}}+\dot{f}(\var)^2}
    % \\ \le &
    \sqrt{f(\var)^2 F_Q[\rho_{\var}]+\dot{f}(\var)^2},
\end{aligned}
\end{equation}
which, after integration and substitution to \eqref{eq:secondstep} gives \eqref{eq:qmi_bound}.

Finally, in the general case where $\rho_{\var}$ can be mixed states, we can find a purification of $\rho_{\var}$ so that the proof above applies.
The quantum Fisher information of $\rho_{\var}$ is equal to the minimum quantum Fisher information of the purification~\cite{escher2011general}.
Let $\ket{\psi_{\var}}$ be a purification that reaches the minimum, i.e., $F_Q[\rho_{\var}]=F_Q[\dyad{\psi_{\var}}]$.
Since $\ket{\psi_{\var}}$ is pure, its Holevo quantity $\chi'$ satisfies,
\begin{equation}
\begin{aligned}
    \chi' \leq &
    \ln\left(\frac{1}{2}\intinf \dd\var\sqrt{F_Q[\dyad{\psi_{\var}}] f(\var)^2+\dot f(\var)^2}\right)
    \\ & - \intinf \dd\var\: p(\var)\ln f(\var),
\end{aligned}
\end{equation}

Meanwhile, the Holevo quantity $\chi'$ is equal to the quantum mutual information between the two subsystems of $\intinf p(\var) \dyad{\var} \otimes \dyad{\psi_{\var}}$ (where $\ket{\var}$ is an orthonormal set), which does not increase under the partial trace operation to $\dyad{\psi_{\var}}$, thus $\chi\le\chi'$ (where $\chi$ is the Holevo information of $\rho$).
Combining all the above, \eqref{eq:qmi_bound} for general mixed states $\rho_{\var}$ is obtained.
$\square$

\end{document}